\documentclass{article}

\usepackage[utf8]{inputenc}
\usepackage{amsmath}
\usepackage{hyperref}

\title{ 
  Polcovar:  
  Software for Computing the Mean and Variance of Subgraph Counts
  in Random Graphs\footnote{The software is available at \url{https://github.com/kunegis/polcovar/}}
}  

\begin{document}

\author{ Jérôme Kunegis }
\maketitle

\begin{abstract}
The mean and variance of the number of appearances of a given subgraph $H$
in an Erdős--Rényi random graph over $n$ nodes are rational polynomials
in $n$ \cite{gp6}.   
We present a piece of software named Polcovar (from \emph{polynomial}
and \emph{covariance}) that computes the exact rational coefficients of
these polynomials in function of the graph $H$.  
\end{abstract}

\section{Introduction}
Large networks can be characterised by the number of times specific
subgraphs appear in them.  For instance, the number of triangles
measures the clustering in a network and the number of wedges
(i.e.\ 2-stars / 2-paths) characterises the inequality of the degree
distribution.  The number of vertices and edges in a graph, too, can be
characterised in this way as the number of times the complete graphs on
one and two nodes appear as subgraphs.  To assess whether a graph
contains many or few subgraphs for its size, the subgraph count must be
compared to the expected subgraph count in random graphs.  To do this,
we must know the distribution of subgraph counts.  Under mild conditions
which are valid for all examples given below, the subgraph count becomes
normal in the large graph limit.  Consequently, the distribution of
subgraph counts is characterised by its mean and average.  Expressions
for the mean and variance of subgraph counts are given in \cite{gp6},
and are always rational polynomials in the number of nodes $n$ of the network.
In this paper, we present Matlab software for computing the exact
coefficients rational
of these polynomials, as their expressions are usually too unwieldy to
be computed by hand. 

\section{Subgraph Counts}
We consider a random graph $G$ on $n$ nodes distributed according to the
Erdős--Rényi model with parameter $p$, i.e., each edge in $G$ exists with
probability $p$ \cite{b569}. Let $H=(W,F)$ be a pattern, i.e. a small graph with
$k=|W|$ vertices and $l=|F|$ edges, and $c_H$ the number of times this pattern
appears as a subgraph of $G$.  
The mean and variance of $c_H$ are then given by the following
expressions~\cite{gp6}:
\begin{align*}
  \mathrm{E}[c_H] &= \frac{n^{\underline{k}}}{|\mathrm{Aut}(H)|} p^{-l} \\
  \mathrm{Var}[c_H] &=
  \sum_J\left(
    \frac{n^{\underline{|V(J)|}}}{|\mathrm{Aut'}(J)|} p^{-|E(J)|} \right)
   - \mathrm{E}[c_H]^2
\end{align*}
where $n^{\underline{i}}$ is the falling factorial\footnote{defined
  as $n^{\underline{i}} = n (n-1) \cdots (n-i+1)$}, 
the sum is over all 
graphs $J$ containing two differently colored copies of $H$ (which might
overlap), 
$\mathrm{Aut}(H)$ is the automorphism group of the graph $H$, 
and $\mathrm{Aut'}(J)$ is the group of automorphisms of $J$ that preserve
the edges of the underlying distinct copies of $H$. 

Although the normal limit for $n \rightarrow \infty$ is only true when the
graph $H$ is \emph{strictly balanced}, the expressions for the mean and
variance are always correct.  Note also that they are true exactly for
any $n$, not just in the large $n$ limit.  

Alternatively, the following expression can be used, which gives the
same result.  It is this expression that we implement in our code.  
\begin{align*}
  \mathrm{Var}[c_H] &=
  - \mathrm{E}[c_H]^2 + \frac{1}{\mathrm{Aut}(H)^2}\sum_{i=0}^k
  \frac{n^{\underline{2k-i}}}{i!(k-i)!^2} 
  \sum_{P,Q} p^{-m(P,Q)}
\end{align*}
where the inner sum is over all pairs of $k$-permutations, and $m(P,Q)$
denotes the number of edges in the overlay of H permuted by $P$ and $H$
permuted by $Q$ which share $i$ nodes.

\section{Special Cases}
For specific small graphs $H$, we get the following exact results. 

\paragraph{Node Count}
Taking $H$ as the graph with one node gives the number of
nodes. Plugging this graph into the general form expression gives
$\mathrm{E}[c_H]=n$ and $\mathrm{Var}[c_H] = 0$.  In other words, the
number of nodes is always exactly $n$, as expected. 

\paragraph{Edge Count}
Edges are always independent of each other and therefore the binomial
approximation for the number of edges $m = c_H$ is exact.
\begin{align*}
  \mathrm{E}[m] &= \frac 1p {n \choose 2} = \frac{n(n-1)}{2p} \\
  \mathrm{Var}[m] &= \frac{n(n-1)}{8} \text{ when $p=1/2$}
\end{align*}
These expressions can be derived both by the general form we gave above,
and by the fact that the number of edges is a binomial distribution. 

\paragraph{Triangle Count}
In a random $n$-graph with parameter $p=1/2$, the number of triangles $t$
has mean and variance given by
\begin{align*}
  \mathrm{E}[t] &= \frac 18 {n \choose 3} \\
  \mathrm{Var}[t] &=
  \frac 1 {128} n^4 - \frac{11}{384}n^3 + \frac 1 {32} n^2 - \frac 1{96} n
\end{align*}

The expressions follow from the general form given below. 

\paragraph{Wedge Count}
The number $s$ of wegdes (i.e., pairs of edges sharing one endpoint,
also known as 2-stars or 2-paths) has
the following distributions when $p=1/2$:

\begin{align*}
  \mathrm{E}[s] &= \frac{n^{\underline{3}}}{8} \\
  \mathrm{Var}[s] &= \frac 18 n^4 - \frac{19}{32}n^3 + \frac {29}{32}n^2
  - \frac{7}{16}n
\end{align*}

The expressions follow from the general form given below. 

\paragraph{Other Patterns}
For the number $q$ of squares we get:
\begin{align*}
  \mathrm{E}[q] &= \frac{1}{128}n^4 -\frac{3}{64}n^3
  +\frac{11}{128}n^2 -\frac{3}{64}n   \\
  \mathrm{Var}[q] &= \frac 1 {512} n^6 - \frac 5 {256} n^5 + \frac
         {161}{2048} n^4 - \frac{163}{1024}n^3 + \frac{327}{2048} n^2 -
         \frac{63}{1024}n 
\end{align*}
For the number $c_H$ of 4-cliques we get: 
\begin{align*}
  \mathrm{E}[c_H] &= \frac{1}{1536}n^4 -\frac{1}{256}n^3
  +\frac{11}{1536}n^2 -\frac{1}{256}n \\
  \mathrm{Var}[c_H] &= \frac{1}{32768}n^6 -\frac{17}{98304}n^5
  +\frac{19}{49152}n^4 -\frac{73}{98304}n^3 +\frac{115}{98304}n^2
  -\frac{11}{16384}n 
\end{align*}

\section{Proof Outline}
A complete proof can be found in \cite{gp6}.  We here outline the proof
as a starting point. The total number of possible subgraphs $H$ in a graph
with $n$ vertices is 
\begin{align*}
  \frac{n^{\underline{k}}}{|\mathrm{Aut}(H)|}.
\end{align*}
Define the random variables $x_i\in \{0,1\}$ to denote the
presence or absence of each possible pattern $i$.  
Then, 
\begin{align*}
  c_H &= \sum_i x_i.
\end{align*}
The expected value of each $x_i$ is given by
\begin{align*}
  \mathrm{E}[x_i] = p^{-l}.
\end{align*}
Thus, the expected value of $c_H$ can be expressed as
\begin{align*}
  \mathrm{E}[c_H] = \mathrm{E}[\sum_i x_i] = \sum_i \mathrm{E}[x_i]
  = \frac{n^{\underline{k}}}{|\mathrm{Aut}(H)|} p^{-l}.
\end{align*}
To compute the variance we exploit the fact that the variance equals the
expected value of the square minus the square of the expected value: 
\begin{align*}
  \mathrm{Var}[c_H] &= \mathrm{E}[c_H^2] - \mathrm{E}[c_H]^2 \\
  &= \mathrm{E}[(\sum_i x_i)(\sum_i x_i)] - \mathrm{E}[c_H]^2 \\
  &= \mathrm{E}[\sum_{ij} x_i x_j] - \mathrm{E}[c_H]^2 \\
  &= \sum_{ij} \mathrm{E}[x_i x_j] - \mathrm{E}[c_H]^2
\end{align*}
Then, each possible pair corresponds to one possible pattern graph $J$,
of which the possible number is
$\frac{n^{\underline{|V(J)|}}}{|\mathrm{Aut'}(J)|}$, and each exists
with independently with probability $p^{-|E(J)|}$.  From this follows
the given expression. 

\section{Extension to Covariances}
The method can be extended to covariances between the count statistics
of different patterns.  As an example:
\begin{align*}
  \mathrm{Cov}[c_{\mathrm{edge}}, c_{\mathrm{triangle}}] &=
  \frac{1}{32}n^3 -\frac{3}{32}n^2 +\frac{1}{16}n
\end{align*}

\section{The Software}
Our code is written in the programming language Matlab, and contains two
entry points, the function \texttt{polcovar\_mu()} that computes the
mean and the function \texttt{polcovar\_sigma()} that computes the
variance or covariance. 

\begin{verbatim}
r = polcovar_mu(H);
r = polcovar_sigma(H1, H2);
\end{verbatim}

The input graphs \texttt{H} must be given as $k \times k$ adjacency matrices.
The function \texttt{polcovar\_sigma()} expects two graphs \texttt{H1} and
\texttt{H2} and returns the covariance of their subgraph counts.  To compute
the variance, pass the same adjacency matrix as both arguments.  All
input matrices must be symmetric 0/1 matrices with zero diagonals. 
All computations are valid for Erdős--Rényi graphs with $p=1/2$. 

The return values are rational polynomials in form of $2 \times (m+1)$
matrices, where $m$ is the degree, coded in the following way:
\begin{align*}
  r &= \left[ \begin{array} {ccccc}
      a_m & a_{m-1} & \cdots & a_1 & a_0 \\
      b_m & b_{m-1} & \cdots & b_1 & b_0 
      \end{array} \right]
\end{align*}
representing the following rational polynomial in $n$:
\begin{align*}
  P_r(n) &= \sum_{i=0}^m  \frac {a_i} {b_i} n ^ i
\end{align*}
All fractions $a_i / b_i$ are returned in simplified form. 

\subsection{Example}
The following example uses Polcovar to compute the mean and standard
deviation of the number of triangles in a random graph with 1,000,000
nodes.

\begin{verbatim}
% Adjacency matrix of a triangle
H = [ 0 1 1; 1 0 1; 1 1 0]

% Compute polynomials
r_mu = polcovar_mu(H)
r_sigma = polcovar_sigma(H, H)

% Evaluate polynomials for a graph with 1,000,000 nodes
n = 1000000
mu = polyval(r_mu(1,:) ./ r_mu(2,:), n)
sigma = polyval(r_sigma(1,:) ./ r_sigma(2,:), n)
sigma_stddev = sqrt(sigma)
\end{verbatim}
This will compute that a random graph with 1,000,000 nodes can be
expected to contain $2.0833 \times 10^{16} \pm 8.8388\times 10^{10}$
triangles.   

\section*{Acknowledgements}
We thank Thomas Sauerwald from the University of Cambridge. 

\bibliographystyle{plain}
\bibliography{polcovar}

\end{document}